\documentclass[a4paper,11pt]{article}
\usepackage{savesym}\savesymbol{widering}\usepackage{yhmath}\restoresymbol{YH}{widering}
\usepackage{pos}
\usepackage{tikz}
\usepackage{slashed}
\usepackage{ulem}
\usepackage{subfigure}
\usepackage{comment}
\usepackage{yhmath}
\usepackage[all]{xy}
\usepackage{multirow}
\usepackage{amsfonts}
\usepackage{float}
\usepackage{mathtools}
\usepackage{xfrac}
\usepackage{wrapfig}
\usepackage{dsfont}
\usepackage{makecell}
\DeclareMathOperator{\llangle}{\big\langle\hspace{-1.2mm}\big\langle\hspace{-.5mm}}
\DeclareMathOperator{\rrangle}{\hspace{-.5mm}\big\rangle\hspace{-1.2mm}\big\rangle}
\newcommand{\bea}{\begin{eqnarray}}
\newcommand{\eea}{\end{eqnarray}}
\newcommand{\beq}{\begin{equation}}
\newcommand{\eeq}{\end{equation}}
\newcommand{\cA}{{\mathcal A}}
\newcommand{\cP}{{\mathcal P}}
\newcommand{\cL}{{\mathcal L}}
\newcommand{\cN}{{\mathcal N}}

\title{Conformal defects and RG flows in ABJM}

\author[a]{Luigi Castiglioni}
\author*[a]{Silvia Penati}
\author[a,b]{Marcia Tenser}
\author[c,d]{Diego Trancanelli}

\affiliation[a]{Dipartimento di Fisica, Universit\`a degli Studi di Milano--Bicocca and INFN, Sezione di Milano--Bicocca, Piazza della Scienza 3, 20126 Milano, Italy}
\affiliation[b]{Universidade Federal de Campina Grande (UFCG),\\ Av. Apr\'igio Veloso 882, 58429-900 Campina Grande, Para\'iba, Brazil}
\affiliation[c]{Dipartimento di Scienze Fisiche, Informatiche e Matematiche, Universit\`a di Modena e Reggio Emilia, \\ via G. Campi 213/A, 41125 Modena, Italy}
\affiliation[d]{INFN Sezione di Bologna, via Irnerio 46, 40126 Bologna, Italy}

\emailAdd{l.castiglioni8@campus.unimib.it}
\emailAdd{silvia.penati@mib.infn.it}
\emailAdd{marciatenser@gmail.com}
\emailAdd{diego.trancanelli@unimore.it} 

\abstract{Defects play a central role in many contexts, from condensed matter to quantum gravity. The situations in which the bulk theory is conformal and the defect inherits part of this symmetry -- the so-called defect conformal field theories (dCFTs) --  have recently received a lot of attention, also thanks to new powerful methods to tackle them, like supersymmetric localization, integrability or the bootstrap. A dCFT may be deformed by turning on marginally relevant operators, which trigger RG flows connecting different fixed points. A natural arena where this phenomenon can be explored are 3-dimensional Chern-Simons theories coupled to matter. These are in fact known to display a plethora of Wilson loops that can be used to define 1-dimensional dCFTs living on their contours. Here we discuss a few examples from an intricate web of RG flows connecting the dCFTs defined on the BPS and non-BPS Wilson loops of ABJM theory. We compute the anomalous dimensions of the deforming operators, establish g-theorems along the flows, and also discuss the role played by cohomological anomalies and framing.}

\FullConference{Proceedings of the Corfu Summer Institute 2024\\ "School and Workshops on Elementary Particle Physics and Gravity" (CORFU2024)\\
12 - 26 May and 25 August - 27 September, 2024, 
Corfu, Greece\\}

\tableofcontents
\begin{document}
\maketitle

%%%%%%%%%%%%%%%%%%%%%%%%%%%

\section{Introduction}

Defects (and boundaries) are of central importance in quantum field theory and find applications in many contexts, from condensed matter to quantum gravity. An interesting class are the defects which are embedded in (super)conformal theories and preserve part of the bulk symmetry. A remarkable example is given by Wilson loops in superconformal theories, which may be seen as defining 1-dimensional defect conformal field theories (dCFTs) living on their contour. One can then study the dCFT, for example by computing correlation functions of local operators inserted on the Wilson loop contour \cite{Cooke:2017qgm}.

Another possibility is to start from a Wilson loop that acts as a UV fixed point, and deform it with a marginally relevant operator. This leads either to another loop operator in the IR or to infinity along runaway directions. The prototypical example is the interpolation, initially proposed in \cite{Polchinski:2011im} and further studied in, for example,  \cite{Beccaria:2017rbe,Beccaria:2018ocq,Correa:2019rdk,Cuomo:2021rkm,Beccaria:2021rmj,Beccaria:2022bcr,Garay:2022szq,Aharony:2022ntz}
\begin{equation}
W^{(\zeta)} = \textrm{Tr} \, \cP \exp \oint \left(i A_\mu \dot x^\mu + \zeta |\dot x| \phi^1\right)d\tau
\end{equation}
between the non-BPS Wilson loop ($\zeta=0$) and the 1/2 BPS Wilson loop ($\zeta = 1$) of 4-dimensional ${\cal N}=4$ super Yang-Mills theory. Here $\phi^{1}$ is one of the scalar fields of the theory, and the interpolation is controlled by a parameter $\zeta$ with $\beta$-function
\begin{equation}
\beta_\zeta = -\frac{\lambda}{8\pi^2}\zeta(1-\zeta^2)+{\cal O}(\lambda^2),
\end{equation}
that acts as a marginally relevant deformation of the non-BPS operator in the UV, triggering a flow toward the supersymmetric operator \cite{Maldacena_1998} in the IR.

It is quite natural to generalize these explorations to 3-dimensional supersymmetric Chern-Simons-matter theories, like ABJM \cite{Aharony:2008ug}, which are known to have a rich moduli space of BPS Wilson loops, see \cite{Drukker:2019bev} for a review. Such operators come in families related by interpolating parameters, that allow to move continuously among representatives preserving varying amounts of supercharges of the theory. One expects then to find a much richer spectrum of flows, given the larger number of Wilson loops that can be defined and serve as fixed points. In this presentation, we discuss some examples of such flows, originally found in \cite{Castiglioni:2022yes,Castiglioni:2023uus}. Some of the flows are purely bosonic, meaning that they only cross Wilson loop operators coupled to the bosonic fields of ABJM, while other flows are mixed, involving both bosons and fermions. Some of the flows are non-BPS, some other flows are `enriched' and preserve some supercharge. This gives an intricate, multi-dimensional space of RG flow trajectories connecting different operators.

This picture has an interpretation in terms of the dCFTs defined on the Wilson loops. In general, deformations driven by local operators $\hat{d}_i$ turn on a defect stress tensor $T_D=\beta_i\,\hat d_i$, where $\beta_i$ is the $\beta$-function of the coupling associated to the $i$-th deformation \cite{Cuomo:2021rkm,Giombi:2022vnz}. This also affects the conservation law of the bulk stress tensor, which gets broken on the defect as
\begin{equation}
    \nabla_{\mu} T^{\mu\nu} =  -\delta_D^{(2)} \left(\dot x^{\nu} \dot T_D + n_i^{\nu} D^i \right) \, ,
\end{equation}
where $\delta_D^{(2)}$ is the Dirac delta localized at the defect, $x^{\mu}$ is the embedding function describing the defect location, $n_i^{\mu}$ is a unit vector normal to the defect and $D^i$ is the displacement operator. 

From $T_D$ one can compute interesting physical quantities, such as the anomalous dimension matrix $\Gamma_i^{\, j}=\frac{\partial \beta^j}{\partial \zeta_i}$ of the $\hat{d}_i$ operators, where $\zeta_i$ is the coupling associated to $i$-th deforming operator. Furthermore, one can establish the validity of the 1-dimensional version of the g-theorem \cite{Affleck:1991tk,Friedan:2003yc,Casini:2016fgb,Casini:2022bsu},
stating that the entropy of a $\zeta$-deformed defect, defined by \cite{Cuomo:2021rkm}
\begin{equation}
    s(\zeta)= \left( 1  + \beta_{\zeta}\frac{\partial}{\partial \zeta} \right) \log \langle W(\zeta) \rangle\,,
\end{equation}
should be a monotonically decreasing function along the RG flows. In particular, since at the fixed points $s$ coincides with the free energy ${\rm g} \equiv \log \langle W(\zeta) \rangle$, this should imply that ${\rm g}_{\rm UV}> {\rm g}_{\rm IR}$, where ${\rm g}_{\rm UV}$ and $ {\rm g}_{\rm IR}$ are the values of the free energy at the UV and the IR fixed points, respectively. From the  mass scaling of $s$ \cite{Cuomo:2021rkm}
\begin{equation}\label{eq:entropyderivative}
    \mu\frac{\partial s }{\partial \mu} = - \int d\tau_{1>2}\,\llangle T_D(\tau_1) T_D(\tau_2) \rrangle (1-\cos(\tau_1 - \tau_2))\,,
\end{equation}
where $\llangle \ldots \rrangle$ indicates insertions along the defect, one concludes that the ${\rm g}$-theorem holds whenever the defect theory is reflection positive (in Euclidean signature) or unitary (in Minkowski), that is whenever $\llangle T_D(\tau_1) T_D(\tau_2) \rrangle > 0$. We will elaborate more on this point in the explicit examples that we are going to describe. 

The plan of this presentation is the following. In section~\ref{sec:WLs} we review the ABJM Wilson loops which define the dCFTs at the fixed points of the flows. Next, we present three examples of RG flows in section~\ref{sec:RGs}. The first example is purely bosonic, the second one is mixed, with both bosonic and fermionic trajectories, and the last one is `enriched'. In section~\ref{sec:dCFTs} we discuss the dCFT interpretation of the flows, computing the anomalous dimension of the deforming operators and establishing g-theorems. The role of cohomological anomalies and their resolution through framing is then discussed in section~\ref{sec:framing}. Finally, we conclude with some generalizations and outlook and sketch in an appendix the perturbative methods we have used. 

%%%%%%%%%%%%%%%%%%%%%%%%%%%

\section{BPS and non-BPS Wilson loops in ABJM}
\label{sec:WLs}

\begin{figure}[!ht]
    \centering
    \includegraphics[width=0.4\textwidth]{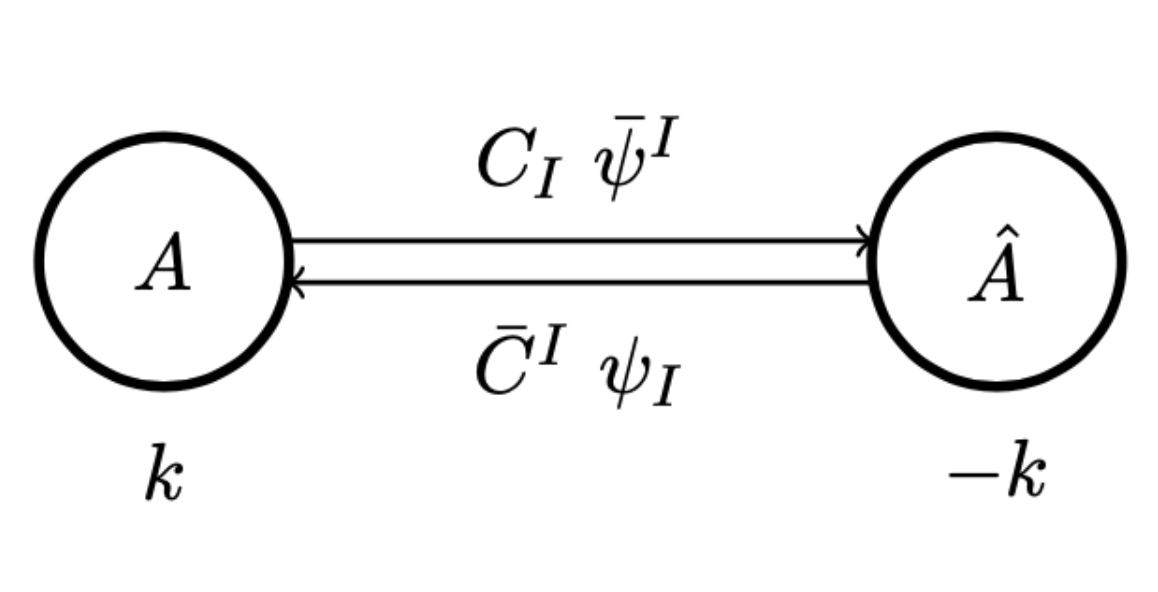}
    \caption{The quiver structure of the ABJM theory.}
    \label{fig:quiver}
\end{figure}

Our bulk theory of interest is 3-dimensional ${\cal N}=6$ super Chern-Simons-matter theory, also known as ABJM \cite{Aharony:2008ug}.\footnote{In section \ref{sec:framing} we also consider the ABJ theory \cite{Aharony:2008gk}, in which the two nodes have gauge groups of different rank, $U(N_1)\times U(N_2)$.} This is a superconformal theory with $OSp(6|4)$ supergroup, whose matter content is described by the quiver diagram in figure \ref{fig:quiver}.
At the two nodes of the quiver there are Chern-Simons gauge fields $A_\mu$ and $\hat A_\mu$ with gauge group $U(N)\times \hat{U}(N)$ and levels $\pm k$, respectively. The matter sector comprises four complex scalars $C_I$ ($\bar C^I$) and four complex spinors $\psi_I$ ($\bar \psi^I$) transforming in the (anti-)bifundamental representation of the gauge group. The index $I=1,\ldots, 4$ is an $SU(4)$ R-symmetry index.

The construction of BPS Wilson loops in ABJM exploits the quiver structure by promoting the usual gauge holonomy to a superconnection $\cL$ in $U(N|N)$
\begin{equation}\label{eq:fermioniWL}
   W=\text{sTr}\, \cP e^{-i\oint \cL d\tau }\,,\qquad 
   \cL=\begin{pmatrix}\cA & \chi_I \bar \psi^I \\ \psi_I \bar\chi^I & \hat\cA \end{pmatrix}\,,
\end{equation}
with 
\begin{equation}\label{eq:gaugeconnection}
\cA = A_\mu \dot{x}^\mu -\frac{2\pi i}{k}|\dot x|M_J^{\ I} C_I \bar{C}^J\,, 
\qquad \hat\cA = \hat{A}_\mu \dot{x}^\mu -\frac{2\pi i}{k}|\dot x| M_J^{\ I}\bar{C}^J C_I\,.
\end{equation}
Here $M_J^{\ I}$ is a scalar coupling matrix and $\chi_I$ ($\bar \chi^I$) are commuting fermionic couplings. We consider the loops to be supported along the circle $x^\mu=(\cos\tau,\sin\tau,0)$ (with $0\le \tau\le 2\pi$) in $\mathbb{R}^3$ and the supertrace in the fundamental representation (of course, different contours and representations are interesting, but beyond the scope of this presentation). The explicit form of the scalar and fermionic couplings determines the amount of supersymmetry preserved by the operator. For example, the case in which $M_J^{\ I}= \textrm{diag}(-1,-1,1,1)$ and $\chi_I=\bar\chi^I=0$ corresponds to the 1/6 BPS bosonic Wilson loop \cite{Drukker:2008zx} preserving 4 out of 24 supercharges, while $M_J^{\ I}= \textrm{diag}(-1,1,1,1)$ and fermionic couplings defined in terms of the projector $\Pi=\frac{1}{2}\left(1+\frac{\gamma \cdot \dot x}{|\dot x|}\right)$ give the 1/2 BPS fermionic Wilson loop of \cite{Drukker:2009hy}.\footnote{For more details on these operators see, for example, the review \cite{Drukker:2019bev}. In particular, there are subtleties about constant shifts in the connections $\cA$ and $\hat \cA$ and alternative formulations in terms of traces, rather than supertraces, which we are being completely glib about here.} More generally, the matter couplings can be deformed by introducing continuous parameters, which play the role of marginally relevant deformations triggering the RG flows discussed in the next section, similarly to what was done in \cite{Polchinski:2011im} for the Maldacena-Wilson loops \cite{Maldacena:1998im} of 4-dimensional $\cN=4$ super Yang-Mills theory.

Before we move on, it is necessary to briefly introduce the set of operators that are going to play a role in the following. These can be classified in terms of the amount of supersymmetry, if any, that they preserve (they all preserve the $SU(1,1)\simeq SL(2,\mathbb{R})$ conformal group on the circle). There are two 1/2 BPS Wilson loops $W_{1/2}^\pm$ preserving 12 supercharges and a $SU(3)$ subgroup of the R-symmetry (thus giving rise to an $SU(1,1|3)$ dCFT). They differ by an overall sign in the scalar coupling matrix $M_J^{\ I}$ and in the fermionic couplings. Next, there is the already mentioned 1/6 BPS bosonic operator $W^\textrm{bos}_{1/6}$, which preserves 4 supercharges and a $SU(2)\times SU(2)$ subgroup of the R-symmetry (it will define an $SU(1,1|1)$ dCFT) and its fermionic counterpart, the 1/6 BPS fermionic Wilson loop $W_{1/6}^\textrm{ferm}$ (which includes, as the name suggests, also a coupling to the fermionic fields). Finally, there is a 1/24 BPS Wilson loop $W_{1/24}$ preserving just one supercharge \cite{Castiglioni:2022yes} and two `ordinary' Wilson loops $W^\pm$, which do not preserve any \cite{Castiglioni:2023uus}.\footnote{See also \cite{Gabai:2022vri,Gabai:2022mya} for a classification of line operators in Chern-Simons theories with matter.} Interestingly, these non-BPS operators do include a coupling to the scalars, through the matrices $M_J^{\ I}= \pm \delta^I_J$, unlike what happens in 4 dimensions, where the ordinary non-BPS Wilson loop of $\cN=4$ super Yang-Mills theory only contains the gauge field holonomy. This novelty in ABJM is due to the fact that scalars in 3 dimensions have scaling dimension 1/2, so that R-symmetry invariant bilinear combinations $\pm C_I\bar C^I$ and $\pm \bar C^I C_I$ can be formed. Under renormalization, these terms mix with the gauge fields and not including them would result in a non-renormalizable dCFT.

%%%%%%%%%%%%%%%%%%%%%%%%%%%

\section{RG flows connecting Wilson loop dCFTs}
\label{sec:RGs}

\begin{figure}[!ht]
    \centering
    \includegraphics[width=0.8\textwidth]{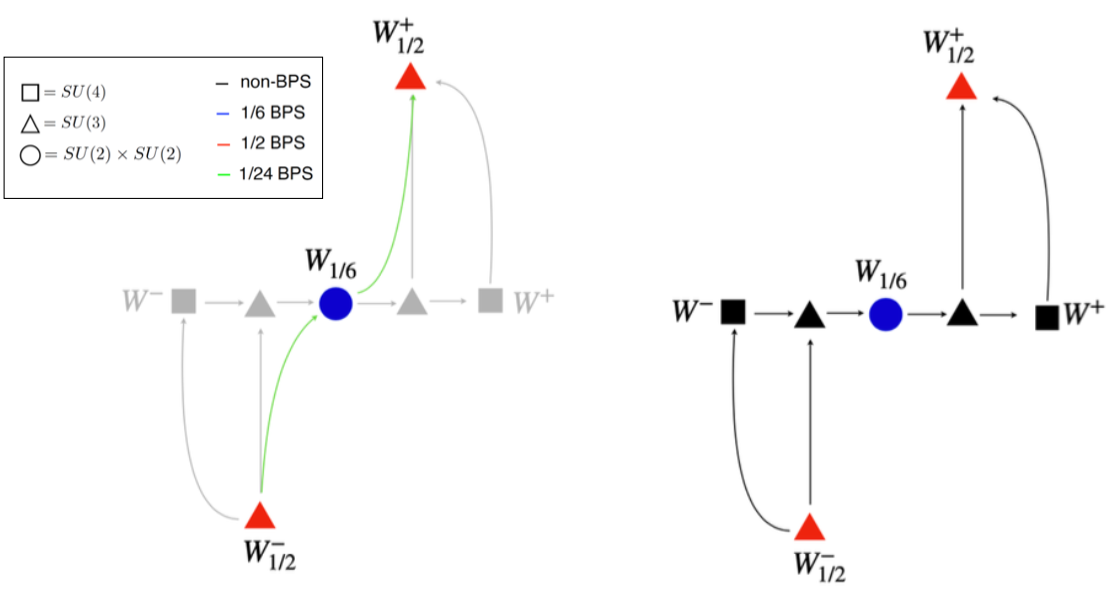}
    \caption{A schematic representation of the flows connecting the dCFTs supported on the Wilson loops. Moving horizontally one spans bosonic operators, vertically the ones with fermionic couplings. The fixed points can be classified by the amount of supersymmetry (colors) and the R-symmetry subgroup (shapes) they preserve. Arrows go from the UV to the IR.}
    \label{fig:flows-schematic}
\end{figure}

As anticipated, one can introduce continuous deformation parameters in the matter couplings $M_J^{\ I}$ and $\chi_I$ ($\bar\chi^I$) in  \eqref{eq:fermioniWL}-\eqref{eq:gaugeconnection}. These act as marginally relevant deformations triggering RG flows among the different dCFTs on the various Wilson loops, see figure \ref{fig:flows-schematic} for a schematic summary.

\begin{figure}[!ht]
    \centering
    \includegraphics[width=0.5\textwidth]{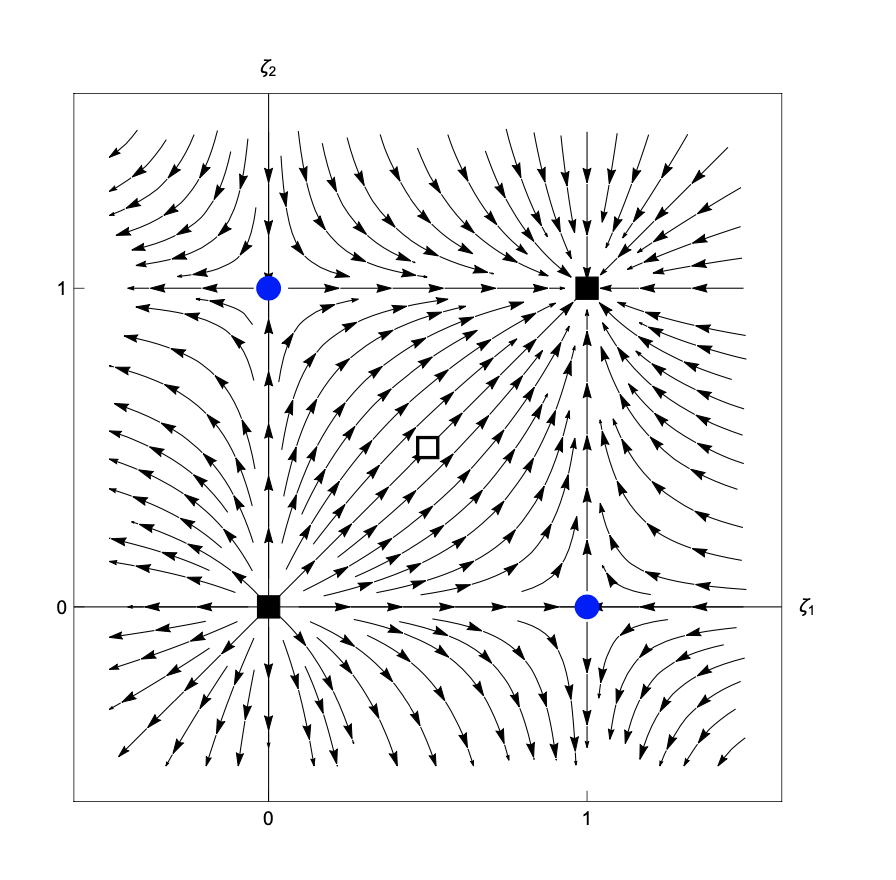}
    \caption{The RG flows in the $(\zeta_1,\zeta_2)$ plane for the bosonic case in \eqref{eq:bosonicdeformations}. The blue dots correspond to the $SU(2) \times SU(2)$ invariant 1/6 BPS bosonic Wilson loops while the black squares are the non-BPS $SU(4)$ invariant ordinary loops $W^\pm$. Note that the operator without any coupling to the scalar fields, the empty square at $\zeta_1=\zeta_2=1/2$, is not a fixed point of the flows. Arrows go from the UV to the IR.}
    \label{fig:bosonicflows}
\end{figure}

As a first example of the procedure followed, let us consider the purely bosonic case (the horizontal flows in figure~\ref{fig:flows-schematic}). For simplicity we restrict the analysis to the first node of the quiver, so that the operators of interest are
\begin{equation}
   W=\textrm{Tr}\, \cP \exp \left[-i\oint \left(A_\mu \dot x^\mu +\frac{2\pi i}{k} C_I\bar C^I-\frac{2\pi i}{k}\Delta M_J^{\ I}C_I \bar C^J  \right)d\tau\right]\,.
\end{equation}
The idea is to start from the non-BPS operator $W^-$ corresponding to $\Delta M_J^{\ I}=0$ and turn on appropriate deformations of the scalar couplings. There are many ways to do that, for example one can request to preserve a $SU(2)\times SU(2)$ subgroup of R-symmetry by choosing
\begin{equation}\label{eq:bosonicdeformations}
\Delta M_J^{\ I}=2\, \textrm{diag}(\zeta_1,\zeta_1,\zeta_2,\zeta_2).
\end{equation}
For $\zeta_1=\zeta_2=1$ one has the other `ordinary' loop $W^+$, while setting $\zeta_1=0$ and $\zeta_2=\pm 1$ (or, equivalently, $\zeta_1=\pm 1$ and $\zeta_2=0$ by the $\mathbb{Z}_2$ symmetry exchanging the two $SU(2)$'s) one recovers the 1/6 BPS bosonic Wilson loops of \cite{Drukker:2008zx}. The $\beta$-functions associated with these parameters can be evaluated in perturbation theory\footnote{See appendix \ref{app:perturbation} for a brief outline of the method employed for these perturbative computations. We stress that our computations, despite being perturbative in the Chern-Simons coupling $N/k$, are exact in the deformation parameters, so that the $\beta$-functions and the general behavior of the RG flows are reliable at any scale.} to lowest order in the coupling constant and give
\begin{equation}\label{eq:betabosonic}
\beta_{\zeta_i}= \frac{N}{k} \zeta_i (\zeta_i-1), \qquad i=1,2.
\end{equation}
The resulting flows are plotted in figure \ref{fig:bosonicflows}. 
The ordinary $W^-$ operator is a UV unstable fixed point, with all arrows pointing outwards, while $W^+$ is an IR stable fixed point. So, even though these two operators only differ by a seemingly trivial overall sign in their scalar couplings, they turn out to be very different under renormalization. Note also that the pure gauge field holonomy (the empty square at the center of the figure) is not a fixed point of the flows. 

\begin{figure}[!ht]
    \centering
    \includegraphics[width=.5\textwidth]{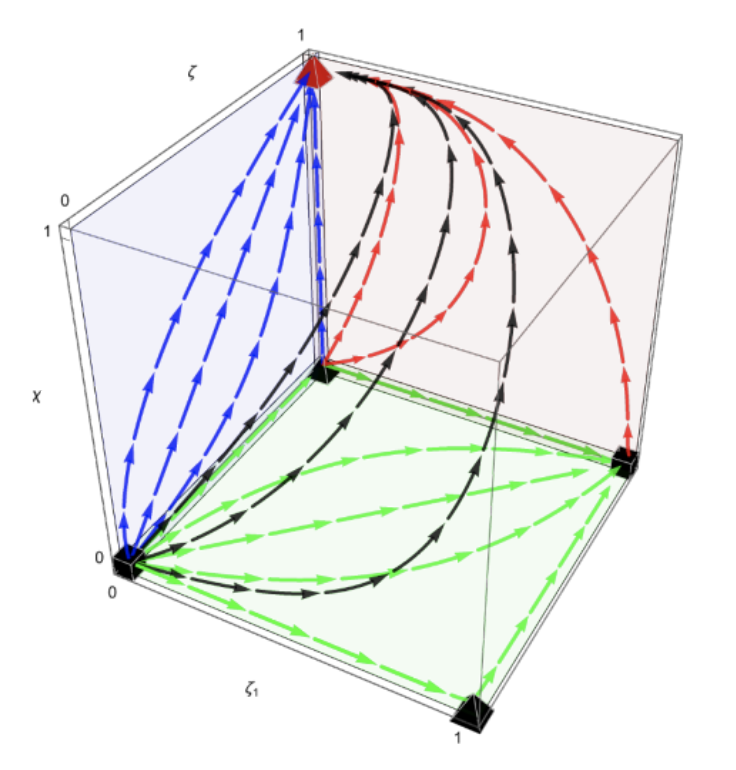}
    \caption{The RG flows in the $(\zeta_1,\zeta,\chi)$ space of the deformations of the 2-node version of $W^-$, which is the operator at the origin and is UV unstable in all directions. The green plane is purely bosonic, while the operator at the top of the cube is the 1/2 BPS Wilson loop $W^+_{1/2}$. Arrows go from the UV to the IR.}
    \label{fig:fermionicflows}
\end{figure}

As a second example, one can now include fermionic couplings, thus enlarging the space of possible RG flow trajectories. Again, there are many possible ways to do this, we refer to \cite{Castiglioni:2023uus} for a large inventory. Here we limit ourselves to starting from the 2-node version of $W^-$ and considering the operator in \eqref{eq:fermioniWL}, with scalar couplings
\begin{equation}
M_J^{\ I} = -\delta_J^I +2\, \textrm{diag}(\zeta_1,\zeta,\zeta,\zeta),
\end{equation}
and fermionic couplings  
\begin{equation}
\chi_I^\alpha=\chi \delta^1_I (e^{i\tau/2}, -ie^{-i\tau/2})^\alpha,\qquad \bar\chi^I_\alpha=\chi \delta^I_1(i e^{-i\tau/2},-e^{i\tau/2})^\textrm{T}_\alpha.
\end{equation}
For $\zeta_1=\zeta=1$ and $\chi=0$ this deformation recovers the other non-BPS Wilson loop, $W^+$. When $\zeta_1=0$ and $\zeta=1$ it gives instead a non-BPS $SU(3)$ invariant bosonic Wilson loop for $\chi=0$ and the 1/2 BPS fermionic operator $W^+_{1/2}$ for $\chi=1$. The $\beta$-functions can be computed as done in the bosonic case at lowest order in the coupling constant (but exactly in the deformations) and read
\begin{eqnarray}
\beta_{\zeta_1}=\frac{N}{k}\zeta_1\left(\zeta_1-1+\chi^2\right),\qquad 
\beta_\zeta = \frac{N}{k}\left(\zeta(\zeta-1+\chi^2)-\chi^2\right),\qquad
\beta_\chi=\frac{N}{k}\chi\left(\chi^2-1\right).
\end{eqnarray}
The operators $W^\pm$, $W_{1/6}^\textrm{bos}$ and $W^+_{1/2}$ are then all fixed points of the flows, which are plotted in figure~\ref{fig:fermionicflows}.

\begin{figure}[!ht]
    \centering
    \includegraphics[width=.5\textwidth]{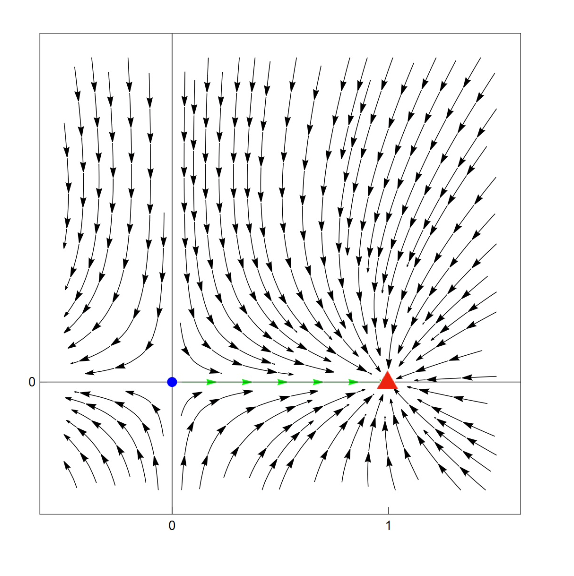}
    \caption{The `enriched' RG flows between the 1/6 BPS Wilson loop $W^\textrm{bos}_{1/6}$ (blue circle) and the 1/2 BPS fermionic operator $W^+_{1/2}$ (red triangle) corresponding to the 1/6 BPS fermionic deformation $W^\textrm{ferm}_{1/6}$ (green line). All other black lines correspond to 1/24 BPS operators. As usual, arrows go from the UV to the IR.}
    \label{fig:enrichedflows}
\end{figure}

In these first two examples, all flows happen to be along non-BPS trajectories, meaning that the Wilson loop operators resulting from the deformation of the fixed points do not preserve any supercharge. It is possible however to consider `enriched' RG flow trajectories, in which all points correspond to operators that preserve at least one supercharge, being then 1/24 BPS \cite{Castiglioni:2022yes}. Specifically, one can deform the 1/6 BPS bosonic Wilson loop $W^\textrm{bos}_{1/6}$ following ideas in \cite{Drukker:2020dvr}. Generic trajectories are 1/24 BPS, while there is a special fermionic deformation in which supersymmetry is enhanced to 1/6 BPS and flows to the 1/2 BPS fermionic Wilson loop, see figure~\ref{fig:enrichedflows}.

%%%%%%%%%%%%%%%%%%%%%%%%%

\section{Defect CFT interpretation}
\label{sec:dCFTs}

The fixed points of the RG flows above are described by 1-dimensional dCFTs living on the Wilson loop contour and interacting with the bulk theory. The RG flows have then the interesting interpretation of flows in the space of 1-dimensional defects triggered by marginally relevant perturbations of the defect.

We discuss this interpretation starting again from the purely bosonic case seen in the previous section, in which one deforms the dCFT defined by $W^-$ and flows uniquely through bosonic theories. Deformation \eqref{eq:bosonicdeformations} preserves $SU(2) \times SU(2)$ symmetry, thus ensuring that the $C_1\bar{C}^1$ and $C_2\bar{C}^2$ operators responsible for the deformation share the same anomalous dimension $\gamma_1$, while $C_3\bar{C}^3$ and $C_4\bar{C}^4$ share the same $\gamma_2$. In principle, there might be mixing between the two pairs of operators, but at the order we are working this is not the case. In fact, the corresponding $\beta$-functions in \eqref{eq:betabosonic} are decoupled and, as a consequence, the anomalous dimension matrix is diagonal, $\Gamma = \text{diag}(\gamma_1,\gamma_1,\gamma_2,\gamma_2)$. 

One can compute the values of $\gamma_1$ and $\gamma_2$ at the four fixed points of figure~\ref{fig:bosonicflows}: at the two black squares, corresponding to $W^-$ (at the origin) and $W^+$, it is $\gamma_1(W^\pm) = \gamma_2(W^\pm) =  \pm N/k$, in agreement with $W^+ \, (W^-)$ being attractive (repulsive). The 1/6 BPS bosonic Wilson loop with coupling $M_J^{\ I} = \text{diag}(-1,-1,1,1)$ (upper blue dot) has instead a negative anomalous dimension, $\gamma_1= -N/k$, and a positive one, $\gamma_2=N/k$. Therefore the deformation with $\zeta_1$ of the $1/6$ BPS Wilson loop is weakly relevant, while the one with $\zeta_2$ is weakly irrelevant. A similar pattern arises for the other 1/6 BPS Wilson loop associated with $M_J^{\ I} = \text{diag}(1,1,-1,-1)$ (lower blue point), simply with the signs of $\gamma_1$ and $\gamma_2$ interchanged. This is in agreement with the saddle point behavior and the directions of the flows shown in figure \ref{fig:bosonicflows}. 

We can easily establish a g-theorem for these RG flows. For simplicity, we consider the $SU(4)$ invariant deformation in which $\zeta_1=\zeta_2\equiv\zeta$, although this can be straightforwardly extended to more general cases.  Defining ${\rm g}_{\rm UV}=\log \langle W^- \rangle$ and ${\rm g}_{\rm IR}=\log \langle W^+ \rangle$, one would like to check whether ${\rm g}_{\rm UV}> {\rm g}_{\rm IR}$. In principle, one could try a direct check by computing these quantities perturbatively. However, this would require a three-loop calculation, as up to two loops $W^\pm$ share the same expectation value. Alternatively, one can use the prescription of \cite{Cuomo:2021rkm}. Inserting $T_D=-\frac{4\pi}{k}\beta_\zeta C_I \bar C^I$ in the scaling equation \eqref{eq:entropyderivative} for the defect entropy, one finds that at one loop ($\tau_{12} \equiv \tau_1 - \tau_2$)
\begin{equation}\label{eq:defectentropy}
    \mu\frac{\partial s}{\partial \mu} = -\frac{16 \pi^2}{k^2}\beta_{\zeta}^2\int d\tau_{1>2} \llangle (C_I\bar C^I)(\tau_1) (C_J\bar C^J)(\tau_2)  \rrangle(1-\cos\tau_{12})=-\frac{4\pi^2 N^2}{k^2}\beta_\zeta^2 <0 \,.
\end{equation}
This means that the defect entropy decreases monotonically along the flow, leading to the expected result.

This discussion can also be reproduced for fermionic operators and fermionic dCFTs. For example, for the $SU(3)$ preserving deformations discussed in the previous section, one can compute the anomalous dimension of the operator
\begin{equation}
  \hat{d} = -\frac{2\pi }{k} \begin{pmatrix} 0 & \chi_1 \bar\psi^1 \\ \psi_1 \bar\chi^1 & 0 \end{pmatrix}
\end{equation}
for the $SU(3)$ invariant bosonic Wilson loop (the black triangle in figure~\ref{fig:fermionicflows}) and the 1/2 BPS fermionic one (the red triangle), confirming that they are negative and positive, respectively. Regarding the g-theorem, a similar computation as in the bosonic case now gives
\begin{align}\label{eq:fermentropy}
    \mu \frac{\partial s}{\partial \mu} = -\frac{\pi^2 N^2}{k} \beta_{\chi}^2<0,
    \end{align}
proving again the monotonically decreasing nature of this quantity.

%%%%%%%%%%%%%%%%%%%%%%%%%%%

\section{Cohomological anomalies and framing}
\label{sec:framing}

The network of RG flows described above connects different fixed points, each of them corresponding to a different dCFT with a non-trivial vacuum $\llangle 0 | 0 \rrangle \equiv \langle W \rangle$. Being this quantity a function of the $(\zeta_i, \chi_I)$ parameters, it acquires different values at different fixed points. This finding is however in contrast with the classical cohomological equivalence \cite{Drukker:2009hy}, which states that, by construction, all BPS Wilson loops differ from the 1/6 BPS bosonic one by a $Q$-exact term, 
\begin{eqnarray}
    W = W^\textrm{bos}_{1/6}+QV\, ,
\end{eqnarray}
where $Q$ is a mutually preserved supercharge and $V$ is a certain operator built out of the fields in the action. If $Q$ is preserved at the quantum level, this implies that for all the Wilson loops we should find $\langle W \rangle = \langle W^\textrm{bos}_{1/6} \rangle$, independently of the values of the parameters. The mismatch between this expectation and our findings reveals  the emergence of a cohomological anomaly at the quantum level, due to the perturbative regularization procedure that breaks superconformal invariance. However, this breaking occurs in a controlled way and can be compensated by choosing a suitable frame, as we now explain. 

An underlying subtlety in the evaluation of Wilson loops in ABJ(M)\footnote{In this section we consider the ABJ case \cite{Aharony:2008gk}, in which the gauge group is $U(N_1)\times U(N_2)$, with $N_1\neq N_2$.} is in fact the one of {\em framing}. This was introduced in pure Chern-Simons theory by Witten in \cite{Witten:1988hf} to regularize a topological anomaly in the partition function caused by gauge fixing. The regularization is achieved by trivializing the tangent bundle and adding a counterterm, introducing in the process a scheme-dependent integer $\mathfrak{f}$. The partition function transforms predictably under a change of scheme, gaining a phase $e^{i\pi\mathfrak{f}\, \text{dim} (G)/12}$ for a gauge group $G$. At the same time, framing restores topological invariance in the expectation value of Wilson loops by regulating short-distance singularities.

In pure Chern-Simons the expectation value of a Wilson loop supported a smooth path $\Gamma$ is given by
\begin{equation}
    \langle W_{\text{CS}} \rangle = \frac{1}{N}\langle \text{Tr}\ \cP e^{-i\int_\Gamma A_{\mu}dx^{\mu}} \rangle\,.
\end{equation}
The perturbative expansion of the operator produces path-ordered integrated $n$-point functions of the form 
\begin{equation}
\int_{\tau_1 > \tau_2 > \dots > \tau_n} \!\! \!\!\!\langle A_{\mu_1}(x_1) A_{\mu_2}(x_2) \ldots A_{\mu_n}(x_n) \rangle \, \dot{x}_1^{\mu_1} \dots \dot{x}_n^{\mu_n} \, d\tau_1 \dots d\tau_n \, , \qquad x_i\equiv x(\tau_i).
\end{equation}
Potential singularities arising from contractions of fields at coincident points can be regularized via a point-splitting procedure, {\it i.e.} by moving the $x_i$'s to auxiliary contours infinitesimally displaced from the original path $\Gamma$:
\begin{equation}\label{eq:framedcontour}
   x^{\mu}_i \to  x^{\mu}_i + \delta\, (i-1)\, n^{\mu}(\tau_i)\,, \qquad |n(\tau_i)|=1\,,
\end{equation}
where $\delta$ is a small deformation parameter and $n^{\mu}(\tau_i)$ are vector fields orthonormal to $\Gamma$ \cite{calugareanu1961}. For example, the one-loop case is the integrated gauge propagator. Letting $x_1$ run on the original contour $\Gamma$ and displacing $x_2$ to the framed contour
\begin{equation}
     \Gamma_\mathfrak{f} :  x_2^{\mu} \to x_2^{\mu}+\delta \, n^{\mu}(\tau_2) \,, \qquad |n(\tau_2)|=1\,,
\end{equation}
one obtains a contribution which is proportional to the Gauss linking integral 
\begin{equation}\label{eq:linking}
    \chi(\Gamma,\Gamma_\mathfrak{f}) = \frac{1}{4\pi}\int_{\Gamma}dx_1^{\mu} \int_{\Gamma_\mathfrak{f}} dx_2^{\nu} \, \epsilon_{\mu\nu\rho} \frac{(x_1-x_2)^{\rho}}{|x_1-x_2|^3}\,,
\end{equation}
a topologically invariant quantity. Sending the deformation parameter $\delta$ to zero does not affect the integral, so that one can define framing as the linking number between $\Gamma$ and $\Gamma_\mathfrak{f}$:
\begin{equation}
\mathfrak{f}\equiv \lim_{\delta\to 0} \chi(\Gamma,\Gamma_\mathfrak{f})\in\mathbb{Z}.
\end{equation}
At higher orders one can check that framing contributions come from diagrams with at least one collapsible gauge propagator and exponentiate, leading for $G=U(N)$ to \cite{Guadagnini:1989am, Alvarez:1991sx}
\begin{equation}\label{eq:pureCS}
    \langle W_{\text{CS}} \rangle_\mathfrak{f} = e^{\frac{i\pi N}{k}\mathfrak{f}}\langle W_{\text{CS}} \rangle_{\mathfrak{f}=0}\,.
\end{equation}

Framing is still relevant in Chern-Simons-matter theories, even though they are non-topological, due to the structure of the gauge propagator. Moreover, matter fields can also contribute to it, leading to non-trivial relations between Wilson loop expectation values at different framing. In ABJ(M) theory in particular, framing plays a crucial role in the discussion of cohomological equivalence, from which one expects, as stated above, that, for example, $\langle W_{1/2} \rangle = \langle W^\textrm{bos}_{1/6} \rangle $. However, this equivalence is not found in explicit perturbative computations \cite{Bianchi:2013zda,Bianchi:2013rma,Griguolo:2013sma}, leading to the necessity to introduce {\it ex post} appropriate phases to compensate for the mismatch \cite{Bianchi:2016yzj}. These phases can be eventually recognized as non-trivial framing contributions. In fact, while ordinary perturbative computations implicitly assume $\mathfrak{f}=0$, supersymmetric localization \cite{Kapustin:2009kz} always computes the Wilson loop expectation value at framing $\mathfrak{f}=1$ \cite{Drukker:2010nc}, where supersymmetry is not broken and all the Wilson loops localize to the same matrix integral. We then conclude that the cohomological anomaly that arises in perturbation theory is totally compensated by the framing phase at $\mathfrak{f}=1$.

More generally, investigating how the results from different framings relate to each other is crucial to compare the results obtained with these different methods and to understand the physical meaning of framing for the defect theory. A direct perturbative computation at generic framing $\mathfrak{f}$ is not an easy task in the presence of matter. For the $1/6$ BPS bosonic Wilson loop a three-loop computation has been carried out in \cite{Bianchi:2016yzj}, showing that the framing-dependent phase \eqref{eq:pureCS} 
 already present in the pure Chern-Simons case gets higher order corrections
\begin{eqnarray}
    \langle W^\textrm{bos}_{1/6} \rangle_{\mathfrak{f}} = e^{\frac{i\pi N_1}{k}\left(1-\frac{\pi^2 N_2^2}{2k^2} \right)\mathfrak{f}} \langle W^\textrm{bos}_{1/6} \rangle_{\mathfrak{f}=0}.
\end{eqnarray}

The presence of fermions further complicates things. It is possible, however, to perform a direct two-loop evaluation of the $1/2$ BPS fermionic Wilson loop at generic framing \cite{Bianchi:2024sod}.\footnote{We thank Marco Bianchi for collaboration on this topic.} The computation relies on carefully isolating framing-dependent contributions in fermionic diagrams and in finding generalizations of the integrand in \eqref{eq:linking} that still evaluate to quantities closely related to $\mathfrak{f}$. The two-loop, non-planar result turns out to be
\begin{equation}\label{eq:W12result1}
    \langle  W_{1/2}\rangle_\mathfrak{f} = e^{\frac{i\pi}{k}(N_1-N_2)\, \mathfrak{f}} \left(1 - \frac{\pi^2}{6k^2} \left(N_1^2 -4N_1N_2+ N_2^2-1\right) \right)+{\cal O}\left(1/k^3\right),
\end{equation}
which is in perfect agreement with the matrix model expectation \cite{Marino:2009jd}, when specializing to $\mathfrak{f}=1$.

%%%%%%%%%%%%%%%%%%%%%%%%%%%

\section{Conclusions}

We have discussed a few representative examples of an intricate web of RG flows connecting the 1-dimensional dCFTs supported on the contour of various Wilson loops of ABJM theory. The fixed points of these flows preserve different amounts of supersymmetry, from zero to half of the supercharges of the bulk theory, and the RG trajectories themselves can either be non-BPS \cite{Castiglioni:2022yes} or `enriched' and preserve some supersymmetry \cite{Castiglioni:2023uus}. The basic idea is to start from certain Wilson loops and deform them with parameters that act as marginally relevant couplings triggering the flows, in the same spirit as what has been done in $\cN=4$ super Yang-Mills theory in \cite{Polchinski:2011im}.

The corresponding $\beta$-functions are computed at first order in the Chern-Simons coupling $N/k$, but exactly in the deformations. A natural question is whether the fixed points we have found are stable under higher order corrections in the bulk coupling constant. One way to answer this question is to study the holographic dual of these flows via a Witten diagram approach \cite{Giombi:2017cqn}, which is something we hope to be able to report on in the near future.

While we have focused here only on circular Wilson loops, other contours are of course interesting. For example, it is possible to consider Wilson loops on cusped lines or on the latitudes of a sphere and proceed to deform them, as done here. This allows to introduce the notion of an interpolating Bremsstrahlung function \cite{Castiglioni:2023tci}, which is computed by taking derivatives of an (interpolating) cusp anomalous dimension with respect to the cusp angle or of the expectation value of the latitude Wilson loop with respect to the latitude angle, generalizing away from the fixed points the cusp/latitude correspondence of \cite{Correa:2012at}.

Other generalizations of this work are also possible, as, for example, considering defects defined on: Wilson loops in higher dimensional representations of $U(N|N)$, along the lines of \cite{Beccaria:2022bcr}; Wilson loops in $\cN=4$ Chern-Simons-matter theories \cite{Drukker:2020dvr,Drukker:2022ywj,Drukker:2022bff,Pozzi:2024xnu}; vortices \cite{Drukker:2008jm,Drukker:2023bip} or higher-dimensional defects. Finally, given the multi-dimensional nature of the RG flows considered here, one could also imagine to apply to our setup the topological approach based on the Conley index proposed in \cite{Gukov:2016tnp} to study spaces of RG flows.

%%%%%%%%%%%%%%%%%%%%%%%%

\section*{Acknowledgements}
We are grateful to Marco Bianchi for collaboration on \cite{Bianchi:2024sod}. This work was supported in part by the INFN grant {\it Gauge and String Theory (GAST)}. DT would like to thank FAPESP’s partial support through the grant 2019/21281-4.

%%%%%%%%%%%%%%%%%%%%%%%%%%%

\appendix
\section{Some details on the perturbation theory}
\label{app:perturbation}

There is a nice way to compute Wilson loops in perturbation theory, which was originally developed for QCD in \cite{Samuel:1978iy, Gervais:1979fv} and then generalized to BPS Wilson loops in ABJM in \cite{Castiglioni:2022yes}. It consists in defining a 1-dimensional auxiliary theory with effective action
\begin{equation} \label{eqn:effS}
   S_\textrm{eff} = S_\textrm{ABJM} +\int d\tau\,  \textrm{Tr} \left(\bar\Psi \left(\partial_{\tau} + i\cL \right)\Psi \right)\,,
\end{equation}
where $\cL$ is the superconnection present in the definition of the Wilson loop under investigation, see \eqref{eq:fermioniWL},  and the $\tau$-integral is taken along the Wilson loop contour. The 1-dimensional Grassmann-odd superfield $\Psi$ has components $z\,(\tilde z$) and $\varphi\,(\tilde \varphi$) that are spinors and  scalars, respectively, in the fundamental representation of the $U(N)$ of the first (second) node of the quiver:
\begin{equation}\label{eq:oddmatrix}
    \Psi = \begin{pmatrix} z & \varphi \\ \tilde \varphi& \tilde z\end{pmatrix}\,, \qquad \bar\Psi = \begin{pmatrix} \bar z & \bar{\tilde\varphi} \\ \bar{\varphi}& \bar{\tilde z}\end{pmatrix}\,.
\end{equation}
The Wilson loop expectation value is then given by the 2-point function of this superfield
\begin{equation}
    \langle W\rangle = \textrm{Tr}\langle \Psi\bar\Psi\rangle \,.
\end{equation}
The right-hand side is computed by integrating over both bulk fields and the superfield $\Psi$, weighted by the action \eqref{eqn:effS}, which can be explicitly written as
\begin{equation}\label{eqn:effectiveaction2node}
   S_{\rm eff} = S_{\rm ABJM} +\int d\tau \Big[ \bar\varphi D_{\tau}\varphi + \bar{\tilde\varphi} \hat D_{\tau} \tilde \varphi  + \bar z D_{\tau} z + \bar{\tilde z} \hat D_{\tau} \tilde z + i\big(\bar{\tilde z} f \varphi + \bar\varphi \bar f \tilde z + \bar{\tilde\varphi} f z+\bar z\bar f \tilde \varphi\big)\Big]\,,
\end{equation}
where $D_{\tau}\equiv\partial_{\tau} +i\cA$ and $\hat D_{\tau}\equiv\partial_{\tau}+i\hat \cA$ include the generalized connections defined in \eqref{eq:fermioniWL} and $f$ and $\bar f$ are its off-diagonal (fermionic) terms. The derivatives give rise to the usual minimal coupling between the 1-dimensional fields and the bulk gauge vectors, plus quartic interactions with bulk scalar bilinears through the scalar coupling matrix $M_J^{\ I}$ appearing in $\cL$. 

\begin{figure}[t]
    \centering
    \subfigure[]{
    \includegraphics[width=0.25\textwidth]{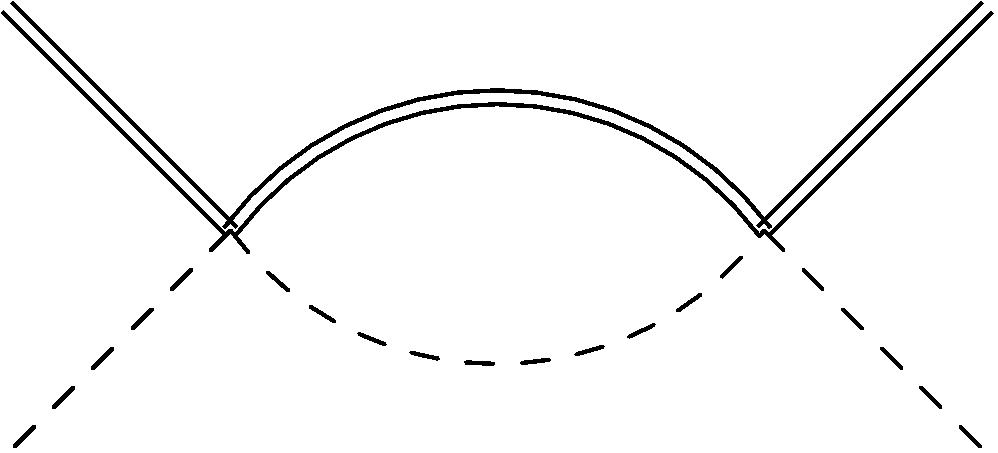}
    \label{subfig:diaga}} \qquad\qquad
    \subfigure[]{
    \includegraphics[width=0.18\textwidth]{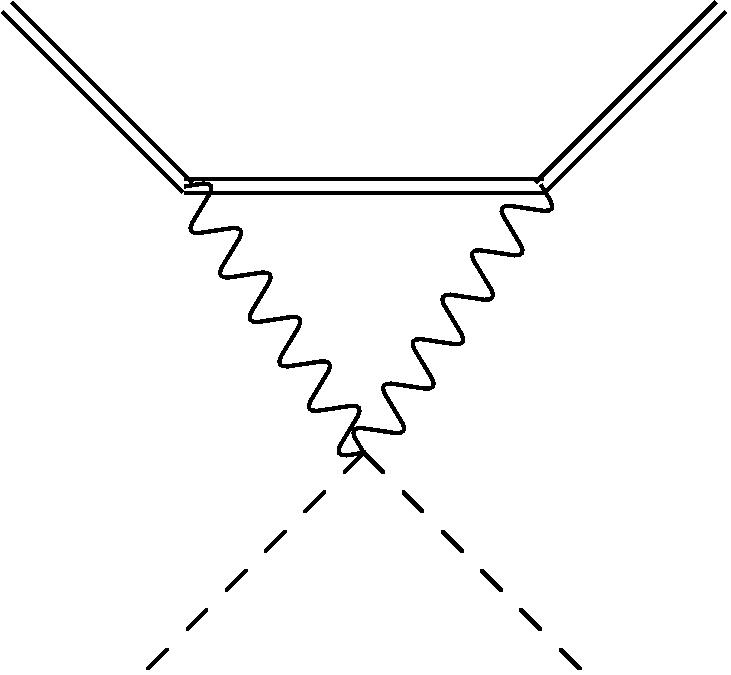}
    \label{subfig:diagb}}  \qquad\qquad
    \subfigure[]{
    \includegraphics[width=0.18\textwidth]{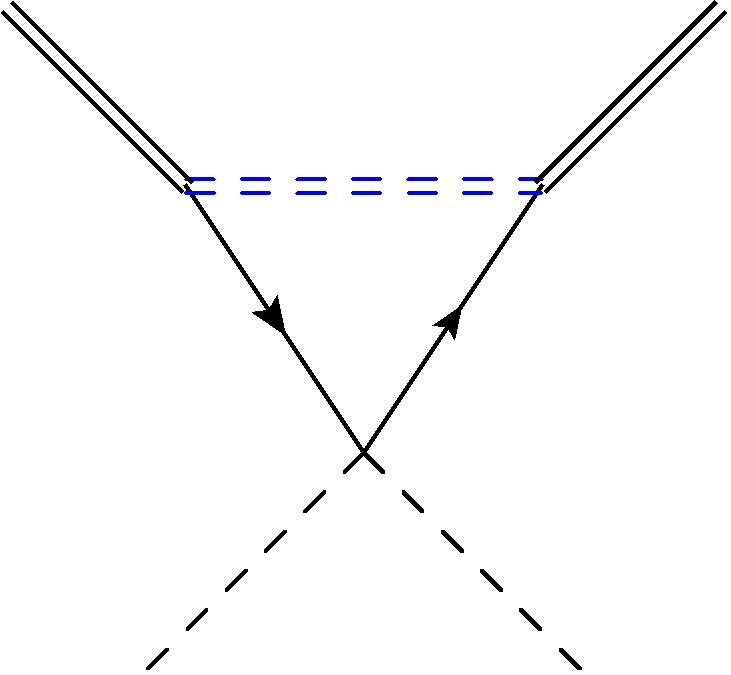}
    \label{subfig:diagc}} 
    \caption{1-loop corrections to the $\bar z C \bar C z$ vertex. Double solid lines represent $z$, double dashed lines $\tilde\varphi$, single dashed lines the scalars $C_I$, solid arrowed lines the fermions $\psi_I$ and wavy lines the gauge field $A_\mu$.}
    \label{fig:diag1}
\end{figure}

To renormalize the theory, for each 1-dimensional field $\phi=(z,\tilde z, \varphi, \tilde\varphi)$ (now seen as bare, $\phi_0$) a corresponding renormalizazion function $\phi=Z^{-1/2}_\phi \phi_0$ is introduced, and the same is done for the deformation parameters. For example, to renormalize the $\zeta_{i=1,2}$ parameters in the bosonic deformation~\eqref{eq:bosonicdeformations}, one defines counterterms through $(\zeta_i)_0 = Z_{\zeta_i} \zeta_i=(1+\delta_{\zeta_i}) \zeta_i$ and considers the scalar vertex $\bar z C \bar C z$, which receives the 1-loop corrections of figure \ref{fig:diag1}. 
Computing these diagrams in dimensional regularization, $d=3-2\epsilon$, with minimal subtraction one finds \cite{Castiglioni:2022yes}
\begin{equation}\label{eqn:z1}
    \delta_{\zeta_i}\zeta_i = \frac{N}{2 k \epsilon}(\zeta_i-1)\zeta_i,
\end{equation}
and, eventually, the $\beta$-functions in \eqref{eq:betabosonic}.

%%%%%%%%%%%%%%%%%%%%%%%%%%%

\bibliographystyle{JHEP}
\bibliography{refs}
\end{document}